\documentclass[12pt]{article}
\usepackage{epsfig}

\def\beq{\begin{equation}}
\def\eeq#1{\label{#1}\end{equation}}
\def\eeqn{\end{equation}}

\def\beqa{\begin{eqnarray}}
\def\eeqa#1{\label{#1}\end{eqnarray}}
\def\eeqan{\end{eqnarray}}

\let\bar=\overbar

\def\Dslash{\not{\hbox{\kern-4pt $D$}}}
\def\dslash{\not{\hbox{\kern-2pt $\del$}}}

\def\msb{{\bar{\ssstyle M \kern -1pt S}}}

\def\Title#1{\begin{center} {\Large {\bf #1} } \end{center}}

\begin{document}

\Title{Small-x Deep Inelastic Scattering via the Pomeron in AdS}

\bigskip\bigskip


\begin{raggedright}  

{\it Richard C. Brower\index{Brower, R.}\\
Department of Physics\\
Boston University\\
Boston MA 02215}
\bigskip


{\it Marko Djuri\'c\index{Djuri\'c, M.}\footnote{speaker}\\
Centro de F\'isica do Porto\\
Universidade do Porto\\
4169-007 Porto, Portugal}
\bigskip


{\it Ina Sar\v{c}evi\'c\index{Sar\v{c}evi\'c, I.}\\
Physics Department and Department of Astronomy and Steward Observatory \\
University of Arizona\\
Tuscon AZ 85721}
\bigskip


{\it Chung-I Tan\index{Tan, C.-I}\\
Department of Physics\\
Brown University\\
Providence RI 02912}
\bigskip
\end{raggedright}

\section{Introduction}

The $AdS/CFT$ correspondence ~\cite{Maldacena:1997re,Witten:1998qj} is a conjectured exact duality between type IIB string theory living in $AdS_5 \times S_5$ spacetime, and $\mathcal{N} = 4 $ SYM, a field theory living on the 4 dimensional boundary of $AdS_5$ . Although it is a conjecture, since its initial formulation overwhelming evidence has emerged to support it. The duality relates states in string theory to operators in the field theory through the relation
\begin{equation}
\left<e^{\int d^4x \phi_i(x)\mathcal{O}_i(x)}\right>_{CFT}\, =\mathcal{Z}_{string}\left[\phi_{i0}(x)\right],\label{AdS/CFT}
\end{equation}
\noindent where $\phi_{i0}(x)=\phi_i(x,z\rightarrow 0).$ The string theory metric is 
\begin{equation}
ds^2\, = \, e^{2A(z)}\, [-dx^+dx^- + dx_\perp dx_\perp + dzdz] + R^2 d^2\Omega_5.\label{metric}
\end{equation}
In the original equivalence, the factor $e^{2A(z)}\, = \, R^2/z^2 $ corresponding to $AdS$ space metric. It is possible to alter this factor in order to deform the $AdS$ geometry and as a consequence get a dual theory which is closer to QCD. For example, if we introduce a sharp cutoff of our space at some value $z=z_0$, and keep the $AdS$ metric for $z<z_0$ we get the so called hard-wall model. One of the key features of the hard-wall model is that the cutoff at $z_0$ introduces a scale in the theory, roughly corresponding to $z_0 \sim \frac{1}{\Lambda_{QCD}}$ with $\Lambda_{QCD}$ the confinement scale. This gives us a very useful phenomenological model which incorporates confinement.

The correspondence works in the planar limit $N_c \rightarrow \infty$ at large, but fixed, value of 't Hooft coupling $\lambda\, =\, g^2N_C$. In the string theory, $\lambda$ emerges as $\lambda = \frac{\alpha'^2}{R^4}$ so we can see a very important feature of the correspondence: large value of $\lambda$ corresponds to a strongly coupled field theory, but a weakly coupled string theory. Thus we have a very useful tool, since we can use a $1/\sqrt{\lambda}$ expansion on the string side to get results in QCD at strong coupling, out of the reach of pQCD.

In particular, we will use this relation to study at strong coupling the process of deep inelastic scattering (DIS), using the $AdS$ Pomeron of Brower, Polchinski, Strassler and Tan ~\cite{Brower:2006ea}, and fit our results to the data collected at HERA ~\cite{:2009wt}. We find a very good agreement with the experiment, obtaining a $\chi^2_{/d.o.f.}\, = \, 1.04$ for our best model. 
%
%
\section{Pomeron in AdS}\label{sec:pomeron}

The Pomeron first emerged in pre QCD days in the 1960's as the leading order exchange in all total cross sections in the Regge limit $s \gg t$. It is an object with the quantum numbers of the vacuum. At high energies, $s\rightarrow \infty$ it can be shown that Pomeron exchange leads to cross sections that rise as 
\begin{equation}
\sigma_{tot} \sim s^{\alpha(0)-1}\label{eq:cross_rise}
\end{equation}
\noindent where $\alpha(0)$ is the Regge intercept of the Pomeron.

In perturbative QCD, the answer to what the bare Pomeron is was first given by Low and Nussinov ~\cite{Low:1975sv,Nussinov:1975mw}. They calculated it as the leading term in the $1/N$ expansion of two gluon exchange, having the topology of a cylinder. The kernel for Pomeron exchange had a cut in the J plane at $j_0 = 1,$ corresponding to a spin 1 exchange. 

Going beyond the leading order, Balitsky, Fadin, Kuraev and Lipatov (BFKL) summed all the diagrams for two gluon exchange to first order in $\lambda = g^2 N_C,$ and {\em all} orders in $(g^2 N_C \log s)^n$, thus giving their namesake equation for the Pomeron exchange kernel. The BFKL kernel was found to have the same symmetries as a $SL(2,\mathcal{C})$ conformal spin chain, which is also the same as the Moebius invariance of scattering amplitudes in string theory. The position of the $j$-plane cut is at $j_0 = 1+ \log (2) \lambda/\pi^2$ recovering the Low-Nussinov result in the $\lambda\rightarrow 0$ limit.

Since we would like to use the $AdS/CFT$ correspondence to study scattering at strong coupling, we need to identify the Pomeron in $AdS$ space. This question was answered by Brower, Polchinski, Strassler and Tan ~\cite{Brower:2006ea}.  They found that, similarly to what happens in the BFKL equation at weak coupling, the Pomeron exchange kernel can be represented as the solution to a $SL(2,\mathcal{C})$ $j$-plane Schrodinger equation on $AdS_3$ space
\begin{equation}
\left[ (-\partial_u^2 - te^{-2u})/2+\sqrt{\lambda}(j-j_0) \right]G_j(t,z,z')=\delta(u-u'),\label{eq:adseq}
\end{equation}
\noindent with $z=e^{-u}$ and the intercept 
\begin{equation}
j_0 = 2 - \frac{2}{\sqrt{\lambda}}.\label{eq:j0string} 
\end{equation}
 The solution to this equation is given by
\begin{equation}
\mathcal{K}(z,z',s,b) = \frac{2(z\,z')^2s}{g_0^2\,R^4}\chi(s,b,z,z'),\label{eq:strongkernel}
\end{equation}
\noindent where
\begin{equation}
\Im{\chi}(s,b,z,z') = \frac{g_0^2}{16\pi}\sqrt{\frac{\rho}{\pi}}\,e^{(1-\rho)\tau}\frac{\xi}{\sinh\xi}\frac{\exp(-\frac{\xi^2}{\rho\tau})}{\tau^{3/2}}\,.\label{eq:chi}
\end{equation}
\noindent Due to conformal invariance, $\chi$ is a function of only two variables:
\begin{eqnarray}
\xi&  = &\log(1+v+\sqrt{v\,(2+v)}) \\
\tau& = &\log(\frac{\rho}{2}z\,z'\,s),
\end{eqnarray}
\noindent where $\rho = 2/\sqrt{\lambda} $ and $v$ is the chordal distance in $AdS_3$, $v\,=\, [(x^\perp-x'^\perp)^2+(z-z')^2]/2zz'.$ In the limit $\tau\gg 1,\,\,\lambda \gg 1$ and $\lambda/\tau \rightarrow 0$, which will be the limit of interest to us in the next section, there is a simple relation between the real and imaginary parts of $\chi$
\begin{equation}
\Re{\chi} \approx \cot(\frac{\pi\,\rho}{2})\Im{\chi}.
\end{equation}
In the infinite coupling limit, $\lambda\rightarrow \infty$ the equation would reduce to graviton exchange, and $j_0\rightarrow 2$ as right for a spin 2 object. Thus the Pomeron is the Regge trajectory of the graviton in $AdS$. The reader can find the details of the derivations of the above results in ~\cite{Brower:2006ea,Brower:2007qh,Brower:2007xg}. A similar $AdS$ analysis was done for the Odderon, another Regge trajectory with the quantum numbers of the vacuum, in ~\cite{Brower:2008cy}. We can notice that the weak and strong coupling Pomeron exchange kernels share a remarkably similar form. At $t=0$, for weak coupling we have
\begin{equation}
\mathcal{K}(k_\perp,k'_\perp,s)\,=\,\frac{s^{j_0}}{\sqrt{4\pi\mathcal{D}\log s}}e^{-(\log k_\perp - \log k'_\perp)^2/4\mathcal{D}\log s},
\end{equation}
\noindent with $j_0 = 1 + \frac{\log 2}{\pi^2}\lambda,\,\,\,\mathcal{D}\,=\,\frac{14\zeta(3)}{\pi}\lambda/4\pi^2$, while at strong coupling
\begin{equation}
\mathcal{K}(z,z',s)\,=\,\frac{s^{j_0}}{\sqrt{4\pi\mathcal{D}\log s}}e^{-(\log z - \log z')^2/4\mathcal{D}\log s},\label{eq:kernelt0}
\end{equation}
\noindent with $j_0 = 2-2/\sqrt{\lambda},\,\,\,\mathcal{D}\,=\,1/2\sqrt{\lambda}$.

\begin{figure}[htb]
\begin{center}
\includegraphics[height=1.85in]{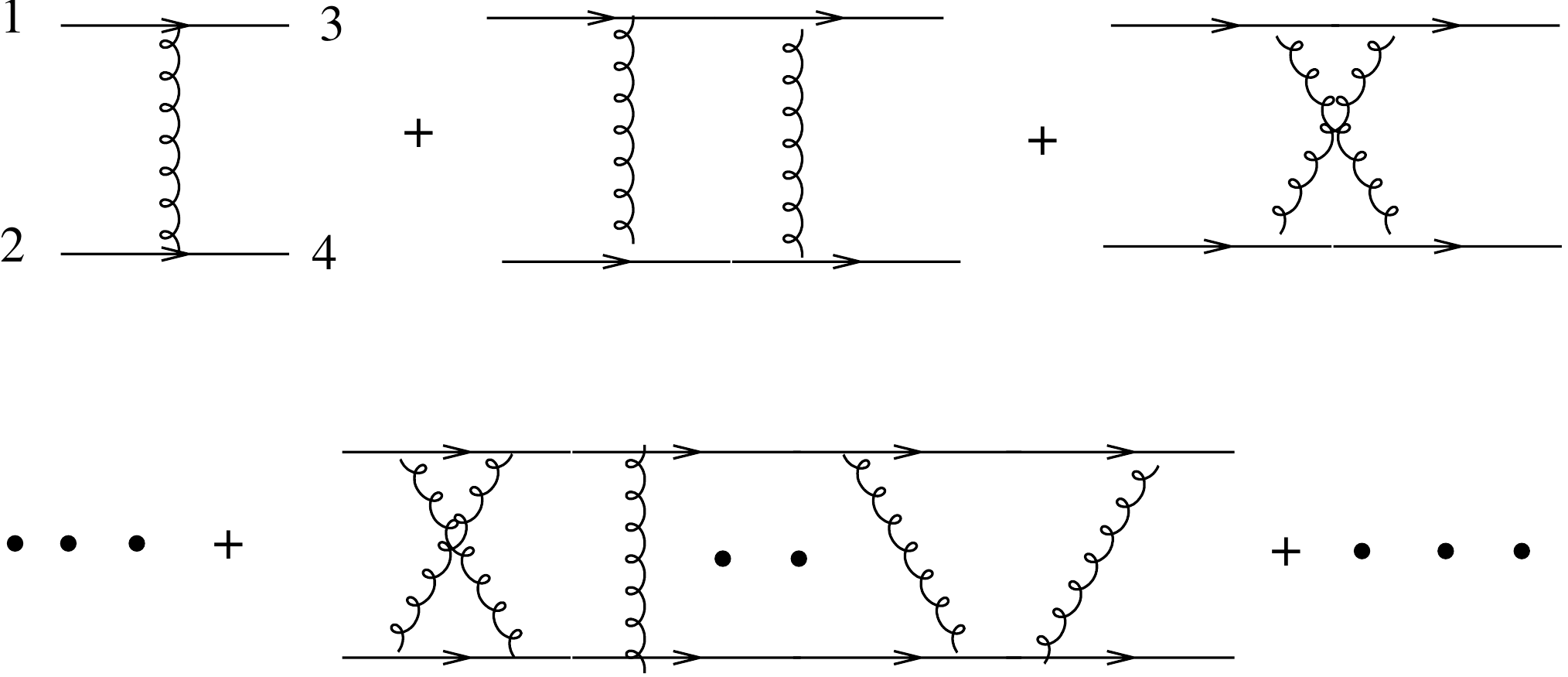}
\caption{The sum of all ladder and cross-ladder diagrams in the eikonal approximation.}
\label{fig:eikonal}
\end{center}
\end{figure}

According to the Froissart bound,
\begin{equation}
\sigma_{tot}\leq \pi c \log^2(\frac{s}{s_0}).\label{eq:froissart}
\end{equation}
Comparing to equation ~\ref{eq:cross_rise} we see that single Pomeron exchange will violate this bound. Therefore eventually effects beyond single Pomeron exchange will start to become important. The eikonal approximation
\begin{equation}
A(s,-\mathbf{q_\perp}^2) = -2is\int d^2b\, e^{-i \mathbf{b_\perp}\cdot\mathbf{q_\perp}}\, (e^{i\chi(s,b)}-1)\label{eq:eikonal_flat}
\end{equation}
\noindent satisfies the unitarity bound, as long as $\Im \chi >0$. We can expand the exponential to get
\begin{equation}
A(s,-\mathbf{q_\perp}^2)=-2is\int d^2b e^{-i \mathbf{b_\perp}\cdot\mathbf{q_\perp}}(i\chi+\frac{(i\chi)^2}{2}+\cdots)\, .
\end{equation}
\noindent This corresponds to all orders, but ignoring all non-linear interactions between the Pomerons. The Feynman diagrams we sum to lead to this approximation can be seen in figure ~\ref{fig:eikonal}.

Similarly the eikonal approximation in AdS space was developed in ~\cite{Brower:2007qh,Brower:2007xg,Cornalba:2006xm,Cornalba:2008qf,Cornalba:2007zb}. In this case the transverse space we integrate over is Euclidean $AdS_3$. The amplitude can be written in the form
\begin{equation}
A(s,-\mathbf{q_\perp}^2) = 2is\int d^2b e^{-i \mathbf{b_\perp}\cdot\mathbf{q_\perp}}\int dzdz'\,P_{13}(z)P_{24}(z')(1-e^{i\chi(s,b,z,z')})\label{eq:eikonal_AdS}
\end{equation}
\noindent where $P_{ij}$ are products of the wavefunctions for the external states, i.e.
\begin{equation}
P_{ij}(z) = (z/R)^2\,\sqrt{g(z)}\,\Phi_i(z)\,\Phi_j(z)\,.
\end{equation}
\section{Deep Inelastic Scattering}

\begin{figure}[htb]
\begin{center}
\includegraphics[height=5in]{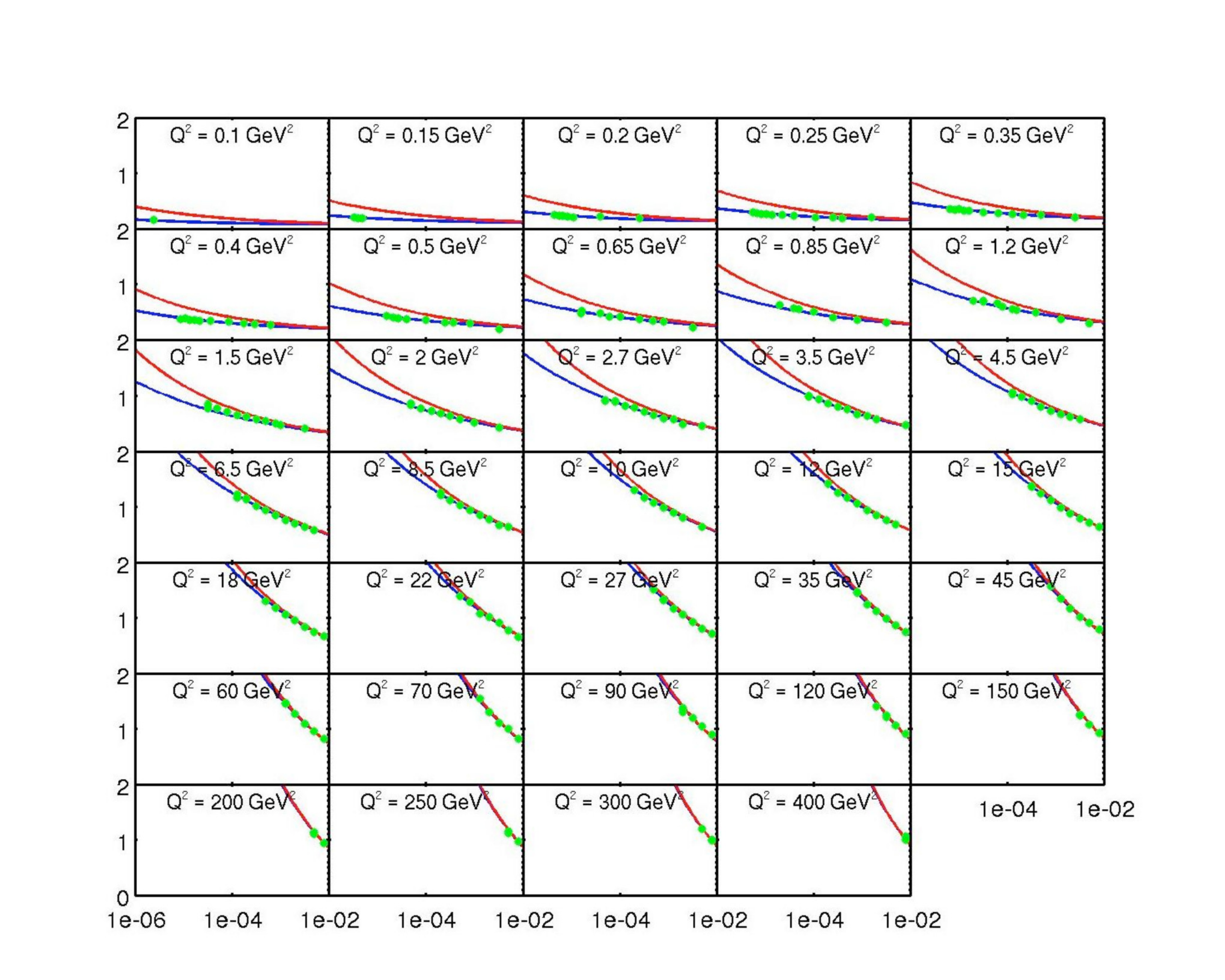}
\caption{The conformal and hard-wall models compared to data.}
\label{fig:1p}
\end{center}
\end{figure}

\begin{figure}[htb]
\begin{center}
\includegraphics[height=3in]{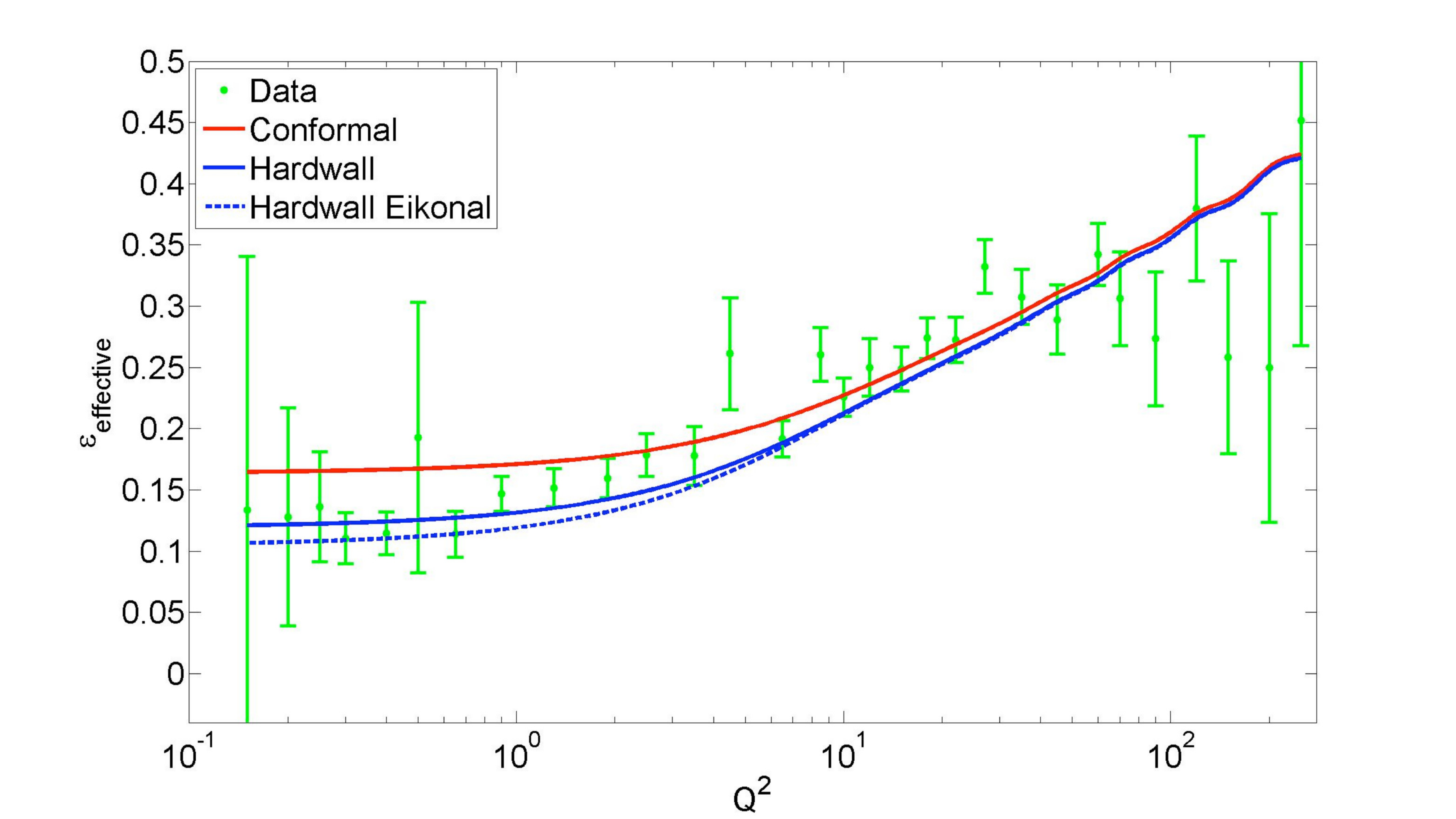}
\caption{$Q^2$-dependence for effective Pomeron intercept, $\alpha_{P}=1+\epsilon_{eff}$.}
\label{fig:eff_pomeron}
\end{center}
\end{figure}

In this section we will present the results found in ~\cite{Brower:2010wf}.  DIS is scattering between an off shell photon, with virtuality $Q^2$, and a proton. We will be interested in the high energy limit, corresponding to small $x=Q^2/s$. In this limit Pomeron exchange dominates, and we are able to apply the $AdS/CFT$ correspondence to find strong coupling results. You can find more details in \cite{Brower:2010wf}, and for some other applications of $AdS/CFT$ to DIS see for example \cite{Cornalba:2010vk,Albacete:2008ze,Kovchegov:2009yj,Levin:2010gc} or \cite{Brower:2010wf} for more complete references. For a similar approach, but using the BFKL Pomeron at weak coupling see \cite{arXiv:1005.0355}. We are interested in calculating the structure function $F_2$
\begin{equation}
F_2(x,Q^2) =\frac{Q^2}{4\pi^2\alpha_{em}}\,\sigma_{tot}(\gamma^*p)\,.\label{eq:strfunction}
\end{equation}
\noindent To calculate the cross section we can relate it to the Pomeron exchange amplitude via the optical theorem, $\sigma=\frac{1}{s}\Im A(s,t=0).$ To get the scattering amplitude, we can apply the results from section ~\ref{sec:pomeron} 
\begin{equation}
\Im A(s,t=0) = \int dz dz'\, P_{13}(z) P_{24}(z') \, \frac{g_0^2 R^4}{2(zz')^2 s}\mathcal{K}(z,z,s),\label{eq:imaginaryA}
\end{equation}
\noindent with $\mathcal{K}$ given in equation (\ref{eq:kernelt0}). Putting it together, using the eikonal approximation (\ref{eq:eikonal_AdS}) we get for $F_2$
\begin{equation}
F_2(x,Q^2) = \frac{ Q^2}{ 2\pi^2} \int d^2b \int dz\,dz' P_{13}(z,Q^2) P_{24}(z')   {\rm Re}\left ( 1-e^{i\chi( s,b,z,z')}\right).  \label{eq:DISeikonal}
\end{equation}
We now need to supply the wave functions $P_{13}(z)$ and $P_{24}(z')$ for the projectile and the target respectively. For the photon, following Polchinski and Strassler ~\cite{Polchinski:2002jw}, we will consider an $R$ boson propagating through the bulk that couples to leptons on the boundary. They calculated the wavefunction in terms of Bessel functions $K_0$ and $K_1$
\begin{equation}
P_{13}(z,Q^2)=\frac{1}{z}(Qz)^{2}(K_{0}^{2}(Qz)+K_{1}^{2}(Qz)), \label{eq:currentF2}
\end{equation}
\noindent We would also need a wavefunction associated to the proton $\phi_p(z)$. For the current analysis, we will assume that the wave function is sharply peaked near the IR boundary $z_0$, with $1/Q'\leq z_0$, with $Q'$ of the order of the proton mass. For simplicity, we will simply  replace $P_{24}$ by a sharp delta-function
\begin{equation}
P_{24}(z')\approx \delta({z'}-1/Q').
\end{equation}
Similarly, for $P_{13}$ which is peaked around $z\simeq 1/Q$, we will replace
\begin{equation}
P_{13}(z)\approx \delta(z-1/Q). \label{eq:p13}
\end{equation}
First we will look at the conformal limit, using single Pomeron exchange. The $b$ space integration of equation (\ref{eq:chi}) can be performed explicitly
\begin{eqnarray}
\int d^{2}b\;  Im\;  \chi (s,b,z,z')
&=&
 \frac{g_{0}^{2} }{16}{\sqrt{\frac{\rho^3}{\pi } } } \; (zz' )\;  e^{(1-\rho) \tau}\frac{\exp(\frac{-(\log z -\log z')^2}{\rho\tau})}{\tau^{1/2}}\,.  
\end{eqnarray}
\noindent Similarly for the aforementioned hard-wall model we would have ~\cite{Brower:2006ea,Brower:2010wf}
\begin{equation}
Im \; \chi_{hw}(s,t=0,z,z')=Im\; \chi_c(\tau,0,z,z')+{\cal F}(z,z',\tau)\; Im\; \chi_c(\tau,0,z,z_0^2/z')\,,\label{eq:Imchi_hw}
\end{equation}
\noindent where
\begin{equation}
\mathcal{F}(z,z',\tau) = 1-4\sqrt{\pi\tau}e^{\eta^2}erfc(\eta)\, , \,\,\,\,\,\,\,  \eta = \frac{-\log(z/z_0)-\log(z'/z0)+4\tau}{\sqrt{4\tau}}\, .\label{eq:F, eta}
\end{equation}
Using these expression in the equation (\ref{eq:DISeikonal}) for $F_2$ staying at single Pomeron exchange level we would have
\begin{eqnarray}
\label{eq:F21P}
F_2(x, Q^2) = \frac{g_{0}^{2} \rho^{3/2}  }{32 \pi^{5/2}} \int &dz dz'P_{13}(z,Q^2)P_{24}(z') {(zz' Q^2)}\; e^{(1-\rho)\tau} \\
&\times \left((\frac{e^{-\frac{\log^{2}z/z'}{\rho\tau}}}{\tau^{1/2}}+ \mathcal{F}(z,z',\tau)\frac{e^{-\frac{\log^{2}zz'/z_{0}^{2}}{\rho\tau}}}{\tau^{1/2}})\right)\,.\nonumber
\end{eqnarray}
The difference between the conformal and hard-wall Pomeron exchange comes in the second term in the above equation. Therefore, modeling confinement using the hard-wall model, the importance of confinement is encoded in the size of the function $\mathcal{F}$. At fixed $z,z'$, $\mathcal{F}$ goes to $1$ as $\tau\rightarrow0$ and to $-1$ as $\tau\rightarrow\infty.$ Hence, at small $x$, ${\cal F}\rightarrow -1$ and  confinement leads to  a partial cancelation for the growth rate. Since ${\cal F}$ is continuous, there will be a region over which ${\cal F} \sim 0$, and, in this region, there is little difference between the hard-wall and the conformal results. Let us also notice the factor
\begin{equation}
e^{(1-\rho)\tau} \sim (\frac{1}{x})^{1-\rho}
\end{equation}
\noindent in equation (\ref{eq:F21P}). Eventually at very small $x$ this factor will dominate, leading the expression (\ref{eq:F21P}) to violate the Froissart bound. 
To compare our results to data, we will fit it to the HERA data set ~\cite{:2009wt}. This is a combined data set of ZEUS and H1 results, and covers 4 orders of magnitude in $Q^2$, from $0.10\, -\, 400\,\,\, GeV^2$. To stay in the small $x$ region where Pomeron exchange dominates, we will only consider points with $x<0.01$. For the conformal single Pomeron exchange, the parameter values we get are
\begin{equation}
\rho=0.774\pm 0.0103, \; g_0^2 =110.13\pm 1.93 , \; Q' = 0.5575\pm 0.0432 \; GeV
\end{equation}
\noindent corresponding to an unacceptably high $\chi^2_{d.o.f.} = 11.7$. If we were to consider a smaller subset of the data, which excludes the low $Q^2$ points, a much better fit would be obtained. Fitting the hard-wall Pomeron to the same full data set from HERA we obtain
\begin{eqnarray}
\rho &= 0.7792\pm 0.0034, \;   g_0^2 =103.14 \pm 1.68 , \; \\
z_{0} &= 4.96 \pm 0.14\, GeV^{-1}, \; Q' =  0.4333\pm 0.0243 \; GeV\nonumber
\end{eqnarray}
\noindent and a much better $\chi^2_{d.o.f.} = 1.07$, after using the sieve procedure ~\cite{Block:2005qm}. In figure ~\ref{fig:1p} you can see a comparison between the conformal and hard-wall models with data. As expected, at large values of $Q^2$ the theory is almost conformal, and the conformal and hard-wall models give the same results. However as we decrease $Q^2$ effects of confinement become increasingly important, and a conformal theory is no longer a good approximation, while the hard-wall model still give very good results. Note also that how well the theory is approximated by the conformal model depends not only on the value of $Q^2$ but the value of $x$ as well. Indeed, even for very large $Q^2$ as we go to very small $x$ we begin to see a difference between the two models. 
As we said, single Pomeron exchange violates the unitarity bound. Therefore we will also do the fits using the eikonal approximation. The conformal eikonal will not improve the results, and will still lead to the violation of the unitarity bound. Therefore we need to look at the hard-wall eikonal. We need the result in $s,t$ space
\begin{eqnarray}
Im\; \chi_{hw}(\tau,t,z,z') &=& Im\; \chi_{hw}(\tau,0,z,z') \\
&+&\frac{ \alpha_0t}{2}  \int_0^\tau d\tau'\int_0^{z_0} d \tilde z \; { \tilde z}^2\; \times \\
 &\times & Im\; \chi_{hw}(\tau',0,z,\tilde z) Im\; \chi_{hw}(\tau-\tau' ,t, \tilde z,z')\nonumber \label{eq:integralequation}
\end{eqnarray}
\noindent Work is underway in evaluating this. We used an approximate treatment which incorporates some of the important features. It can be shown ~\cite{Brower:2007xg,Brower:2010wf} that at large $b$ the eikonal for the hard-wall model has a cut-off
\begin{equation}
Im\; \chi_{hw}(\tau ,b,z,z')\sim \exp[-  m_1 b  - (m_0-  m_1)^2\;  b^2 / 4 \rho \tau ]
\end{equation}
\noindent where $m_1$ and $m_0$ are  solutions of
\begin{equation}
\partial_z(z^2J_0(mz))\; |_{z=z_0} =0
\end{equation}
\noindent and
\begin{equation}
\partial_z(z^2J_2(mz))\; |_{z=z_0} =0
\end{equation}
\noindent respectively. For $b$-small, we shall take $Im \; \chi_{hw}(\tau,b,z,z')$ to be of the form
\begin{equation}
Im \;\chi^{(0)}_{hw}(\tau,b,z,z')\sim  Im \;\chi_{c}(\tau ,b,z,z') + {\cal F}(\tau ,z,z')Im \; \chi_{c}(\tau,b,z,z_0^2/z')
\end{equation}
\noindent We therefore adopt the following simple ansatz
\begin{equation}
Im \; \chi_{hw}(\tau,b,z,z') = C(\tau,z,z')D(\tau ,b)  Im\;  \chi^{(0)}_{hw}(\tau,b,z,z')
\end{equation}
\noindent where
\begin{equation}
D(\tau ,b) =   \left\lbrace
\begin{array}{ll}
1\;, & b< z_0\\
\frac{ \exp[-  m_1 b  -  (m_0-m_1)^2\;  b^2 / 4 \rho \tau ]}{\exp[-  m_1 z_0  - ( m_0-m_1)^2\;  z_0^2 / 4 \rho \tau ]}    \;, & b > z_0 \\
\end{array}
\right .
\label{eq:damping}
\end{equation}
\noindent $C(\tau,z,z')$ is an overall normalization constant which we can fix by requiring our result to recover the $t=0$ result. Fitting this expression we get the parameters: 
\begin{eqnarray}
\rho &= 0.7833\pm 0.0035, \;   g_0^2 =104.81 \pm 1.41 , \; \nonumber \\
z_{0} &= 6.04\pm 0.15 \, GeV^{-1}, \;  Q' =0.4439\pm 0.0177 \; GeV
\end{eqnarray}
\noindent The parameters here are very similar to what we had from the hard-wall single Pomeron exchange, and the corresponding $\chi^2_{d.o.f} = 1.04$ is slightly improved. We can also fit the data to `effective Pomerons', by fixing $Q^2$, and then fitting $F_2(x,Q^2)\sim (1/x)^{\epsilon_{eff}}$. In figure ~\ref{fig:eff_pomeron} we see that using the Pomeron in $AdS$ space we interpolate between Pomerons at very different intercepts, and therefore unite the `soft' and `hard' Pomerons.
\begin{figure}[htb]
\begin{center}
\includegraphics[height=3.5in]{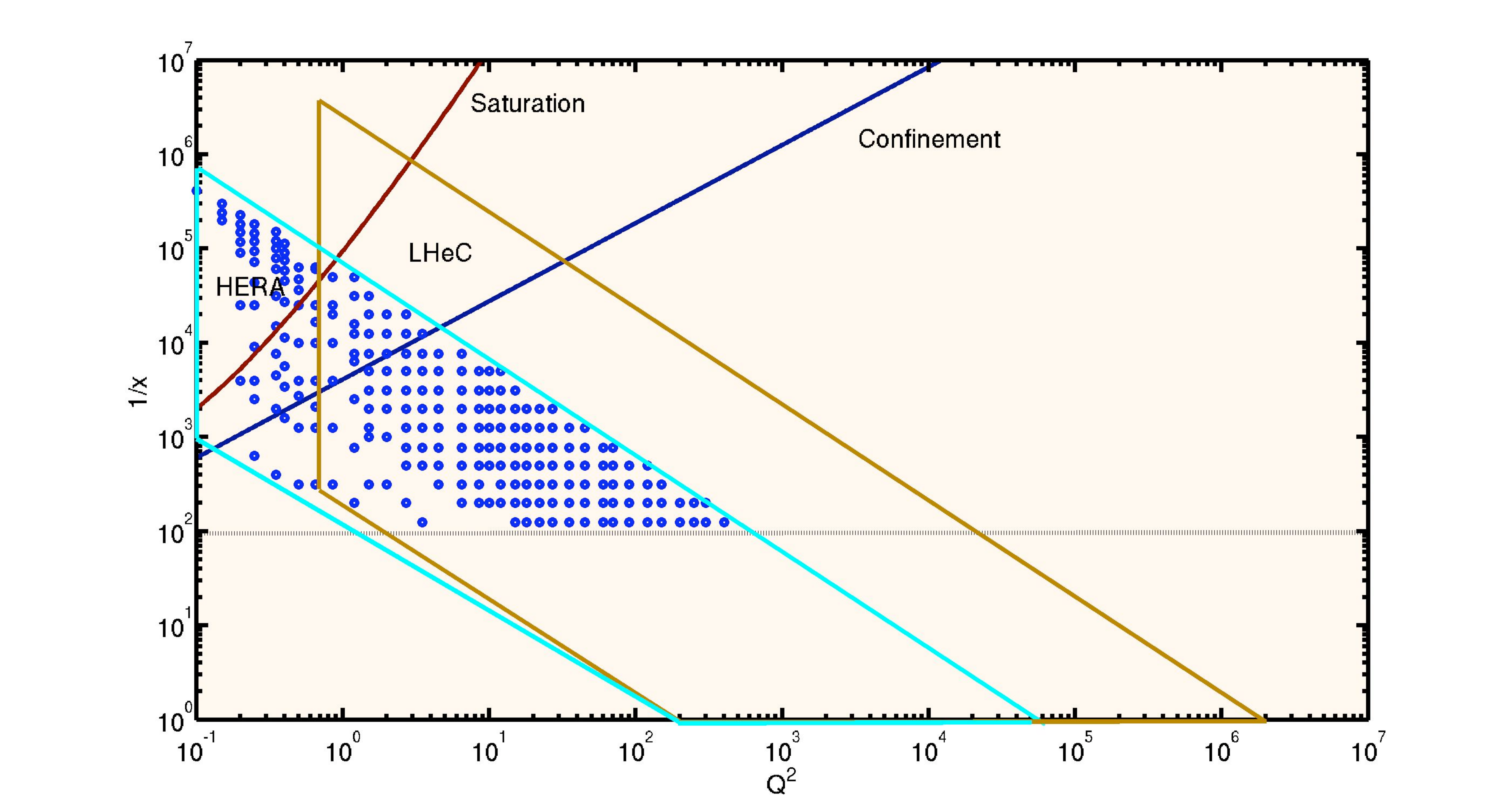}
\caption{The relative importance of saturation and confinement at LHC energies.}
\label{fig:confinement}
\end{center}
\end{figure}
\section{Other Processes}

\begin{figure}[htb]
\begin{center}
\includegraphics[height=3.75in]{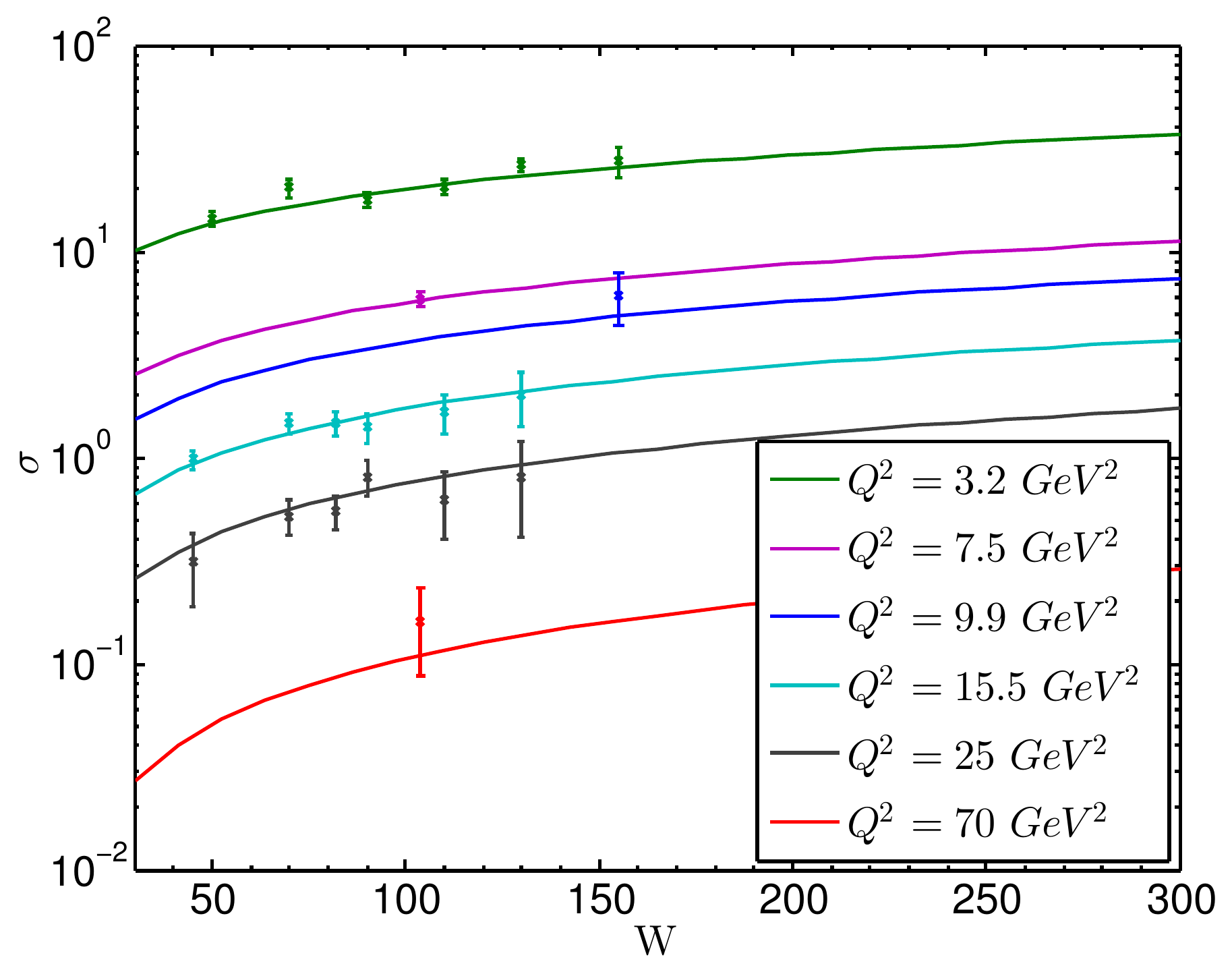}
\caption{Pomeron exchange applied to DVCS, for different values of $Q^2$ ~\cite{DVCS}.}
\label{fig:DVCS}
\end{center}
\end{figure}

Similar methods, using equation (\ref{eq:eikonal_AdS}), can be applied to other scattering processes, for example proton-proton total cross sections, by supplying the appropriate wave functions. It is also possible to extend this framework to $2\rightarrow 3$ scattering, and constrain the values of the parameters using the results presented here. This could then be applied to double diffractive Higgs production at the LHC ~\cite{double_higgs}. Another interesting application would be to deeply virtual compton scattering (DVCS). In this process we have the Compton scattering of a virtual photon and a proton. It is similar to DIS, but it is an exclusive process. Therefore we cannot use the optical theorem for single Pomeron exchange, and we need the full amplitude. Also, more care is needed in evaluating the wavefunction of the incoming and outgoing photons, as they will now depend on the momentum transfer. Work is under way in using the techniques presented above to evaluate this process ~\cite{DVCS}, but preliminary results show that the $AdS$ BPST Pomeron exchange can give good agreement with experiment here as well (figure ~\ref{fig:DVCS}).  
\section{Conclusions}

We have seen that the application of the $AdS/CFT$ correspondence through the BPST Pomeron exchange can give very good fits to data, with few phenomenological parameters. It is important to note that our results point that to have a realistic theory it is crucial to include the effects of confinement. In figure ~\ref{fig:confinement} we see that effects of confinement become important before the effects of saturation. It will be interesting to explore if this feature is a consequence of our model or model independent. Also interesting will be the study of multiple scattering processes to constrain the $AdS$ building blocks which can then be applied to double diffractive production of the Higgs boson (also known as the SX boson), applicable to current LHC experiments.

\paragraph*{Acknowledgments:}

The work of R. C. B.  was supported by the Department of Energy under
contract~DE-FG02-91ER40676, that of IS by the Department of Energy under contracts DEFG02-04ER41319 and DE-FG02-04ER41298 and that of  C.-IT.  by the
Department of Energy under contract~DE-FG02-91ER40688, Task-A. Centro de F\'isica do Porto is partially funded by FCT and the work of M.D. is partially supported by grants PTDC/FIS/099293/2008 and CERN/FP/116358/2010 and by the FCT/Marie Curie Welcome II project.






 
\end{document}